\documentclass[prl,showpacs,twocolumn,amsmath,superscriptaddress]{revtex4}

\usepackage{graphicx}
\usepackage{dcolumn}
\usepackage{bm}

\usepackage{epstopdf}

\begin{document}

\title{Hydroxyl vacancies in single-walled aluminosilicate and aluminogermanate nanotubes}

\author{G. Teobaldi*}
\affiliation{Surface Science Research Centre,
Department of Chemistry, University of Liverpool, L69 3BX Liverpool, UK}
\email{g.teobaldi@liv.ac.uk}
\author{N. S. Beglitis}
\affiliation{London Centre for Nanotechnology,
17-19 Gordon Street, London WC1H 0AK, UK
and Department of Physics and Astronomy,
University College London, Gower Street, WC1E 6BT, UK}
\author{A. J. Fisher}
\affiliation{London Centre for Nanotechnology,
17-19 Gordon Street, London WC1H 0AK, UK
and Department of Physics and Astronomy,
University College London, Gower Street, WC1E 6BT, UK}
\author{F. Zerbetto}
\affiliation{Dipartimento di Chimica "G. Ciamician",
Universit{\`a} degli Studi di Bologna,
via Selmi 2, 40126 Bologna, Italy}
\author{W. A. Hofer}
\affiliation{Surface Science Research Centre,
Department of Chemistry, University of Liverpool, L69 3BX Liverpool, UK}

\date{\today}

\begin{abstract}
We report the first theoretical study of hydroxyl vacancies in aluminosilicate and aluminogermanate
single-walled metal-oxide nanotubes. The defects are modeled on both sides of the tube walls and
lead to occupied and empty states in the band gap  which are highly localized both in energy and in real space.
We find different magnetization states depending on both the chemical composition and the specific side
with respect to the tube cavity.
The defect-induced perturbations to the pristine electronic structure are related to
the electrostatic polarization across the tube walls and the ensuing change in
Br{\o}nsted acid-base reactivity.
Finally, the capacity to counterbalance local charge accumulations,
a characteristic feature of these systems, is discussed in view of
their potential application as insulating coatings for one-dimensional conducting nanodevices.

\end{abstract}


\maketitle

Since the discovery of carbon nanotubes by Ijima in 1991~\cite{1},
the field of carbon-based and inorganic nanotube materials has grown considerably
due to their potential applications in electronics, photonics, chemical separation,
(photo)catalysis and biotechnology~\cite{2}.
While carbon-nanotubes and their inorganic counterparts such as BN, $\mathrm{WS_2}$  and $\mathrm{MoS_2}$
are routinely produced by electric-arc discharge, chemical vapor deposition,
or laser ablation~\cite{2}, new low-temperature solution-phase chemistry routes
have recently been proposed for semiconducting single-walled aluminosilicate (Al-Si)
and aluminogermanate (Al-Ge) nanotubes (see Fig. 1 and~\cite{3}).
The reported extremely high experimental control in terms of dimensions
and monodispersity of these nanostructures~\cite{3},
together with the potentially huge range of tunable properties 
via chemical functionalization and substitutional doping~\cite{5},
make both Al-Ge and Al-Si attractive candidates
as large-storage chemical devices~\cite{3}, artificial ion-channel systems~\cite{2},
and insulating coatings for conducting cores,
which could make it possible to enforce one-dimensional anisotropic conductivity at the nanoscale~\cite{ZEOM}.
Al-Si and Al-Ge nanotubes are structurally analogous
to the naturally occurring hydrous-aluminosilicate Imogolite~\cite{9}.
Its walls consist of a single layer of octahedrally coordinated aluminium hidroxide (gibbsite),
with tetrahedral silanol (Si-OH) groups attached at the inner side of the tube (see Fig. 1).
From a compositional point of view,
the only difference between Al-Si and Al-Ge tubes is the substitution of silanol groups
with germanol (Ge-OH) moieties. The resulting chemical formula of the unit cell
is $\mathrm{(Al_2SiO_7H_4)_N}$ and $\mathrm{(Al_2GeO_7H_4)_N}$ for the Al-Si and Al-Ge tubes, respectively.
$N$ refers to the number of radially inequivalent aluminum atoms along the nanotube circumference,
necessarily an even number for symmetry reasons~\cite{10}.
Regardless of the specific route adopted for their synthesis,
both Al-Si and Al-Ge tubes are achiral and analogous to
zig-zag $\mathrm{(n,0)}$ semiconducting carbon-based nanotubes~\cite{CNT}. 
Following a recent Density Functional Theory (DFT) study of Imogolite-based nanotubes~\cite{MEX},
which opened up the possibility of first-principle studies for these systems,
as well as previous simulations of neutral paramagnetic defects
in zeolites~\cite{Catlow}
we address the effects of the simplest defects, i.e.
neutral hydroxyl (-OH) vacancies, on Al-Ge and Al-Si in terms
of both electronic structure changes and ensuing modifications in the global insulating properties.

\begin{figure}[htbp]
\begin{center}
\includegraphics[width=\columnwidth]{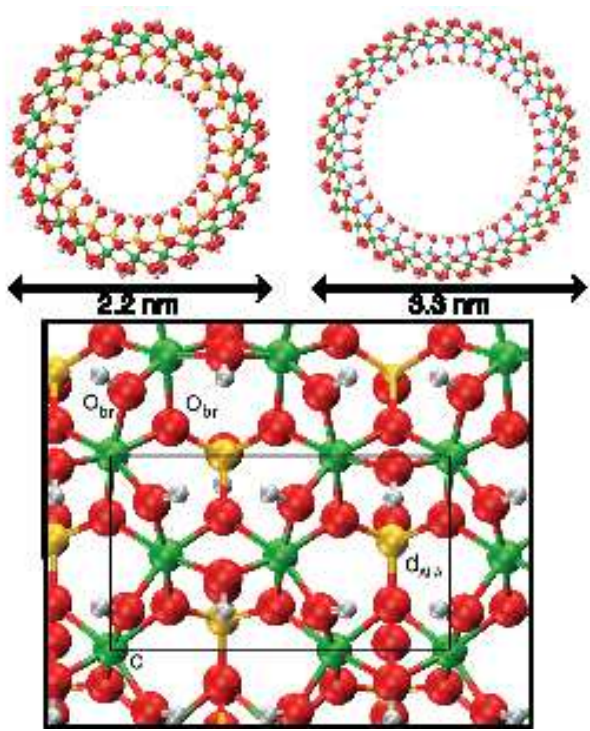}
\end{center}
\caption{(Color online) Optimized geometrical structure of $\mathrm{(Al_2SiO_7H_4)_{24}}$ (left)
and $(\mathrm{Al_2GeO_7H_4)_{36}}$ (right) based nanotubes.
The single-wall structural motif (bottom) is displayed together with
the zig-zag periodic unit of size $c, d_{Al-Al}$ along the nanotube axis and circumference.
O: red, H: gray, Al: green, Si: yellow, Ge: cyan.}
\label{fig1}
\end{figure}

DFT simulations were performed within a plane-wave ultrasoft pseudopotential framework
as currently implemented in the VASP code~\cite{VASP}.
On the basis of previous successful results for Al-Si(Ge) based systems~\cite{MEX},
aluminosilicate mesostructures~\cite{ALSIclust} and aluminium oxide thin films~\cite{Al-Kresse},
exchange and correlation were approximated
at the semi-local generalized gradient-corrected PW91-GGA level~\cite{PW91}.
Hybrid B3LYP~\cite{B3LYP} periodic calculations were carried out with
the Gaussian program~\cite{Gaussian}, adopting a CEP-121G triple-zeta basis set~\cite{CEP-121G}.
In the spirit of ~\cite{3} (and at odds with ~\cite{MEX}) we considered and optimized
(see Supporting Information, SI) the repeat unit cell ($c$ in Fig. 1)
for Al-Si (Al-Ge) systems made up of $N=24$ ($N=36$) gibbsite units.
The optimized c value of 8.68$\pm$0.01 $\mathrm{\AA}$ (8.78$\pm$0.01 $\mathrm{\AA}$)
is slightly larger than experimental X-ray diffraction (XRD) data
for Al-Si (Al-Ge) thin films i.e. 8.51 $\mathrm{\AA}$~\cite{3}.
The deviations are due to the well-known overestimation of aluminosilicate bond lengths by the semi-local
density functional approximation~\cite{ALSIclust,Gale-GGA,Clay-GGA}.
We also note that the agreement between the calculated values of 8.68 $\mathrm{\AA}$ for $N=24$ 
considered here and the reported 8.62 $\mathrm{\AA}$ for N=20 
(equivalent to $\mathrm{N_u}$=10 in ~\cite{MEX}) confirms that
the size of the unit cell depends only weakly on $N$~\cite{MEX}.
The optimized diameter of 23.15$\pm$0.05 (33.1$\pm$0.04) $\mathrm{\AA}$ is
in good agreement with the reported experimental value of
$\sim$22 ($\sim$33) $\mathrm{\AA}$ for Al-Si (Al-Ge) ~\cite{3}.
By independently optimizing the bi-dimensional (2D) analogues of Al-Si and Al-Ge (see SI),
the bending energy was evaluated to be \-0.45 eV/$N$ and \-0.19 eV/$N$ for Al-Si and Al-Ge tubes respectively. 
These results further confirm that the tube bending is a consequence of
minimizing the strain from the mismatch between the stronger Si(Ge)-O and
weaker Al-O bonds (see SI and ~\cite{3,9}) in that the more bent Al-Si
is more stabilized than Al-Ge by the bending process.
We also note that on the basis of the calculated smaller Al-Ge formation energy,
it is tempting to link these results with the experimental evidence
of longer ($\mathrm{\sim}$100 nm) Al-Si tubes and
shorter ($\mathrm{\sim}$20 nm) Al-Ge tubes~\cite{3}.

\begin{figure}[htbp]
\begin{center}
\includegraphics[width=\columnwidth]{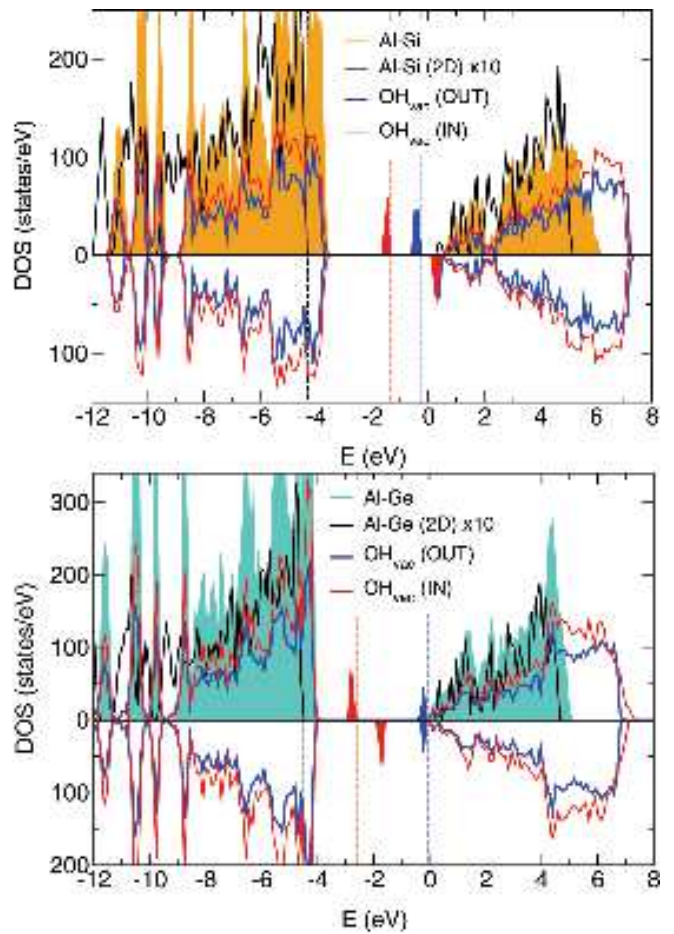}
\end{center}
\caption{(Color online) Total Density of States (DOS) for defect-free Al-Si and Al-Ge,
their 2D analogues, and in the presence of one $\mathrm{OH_{vac}}$ both on the outer (OUT)
and the inner (IN) surface of the tubes.
Calculated Fermi energies are displayed as dotted line with the same color labeling as for the DOS.
2D and band gap defect states (filled) have been increased by a factor 10 for clarity.}
\label{fig2}
\end{figure}

The calculated Density of States (DOS) for Al-Ge, Al-Si and the corresponding 2D sheets are reported in Fig. 2.
The calculated band gaps are 4.1 eV and 3.9 eV for Al-Si and Al-Ge respectively.
Besides being in close agreement with the experimentally reported value of 3.6 eV for Al-Ge~\cite{3},
these values are in accordance also with simulations of
aluminosilicate clusters ($\sim$4 eV~\cite{ALSIclust}) and
aluminum oxide thin films ($\sim$4 eV~\cite{Al-Kresse})
obtained using the same PW91 functional.
Finally, we note that all data match also the calculated band gaps recently reported for (N=20) Al-Si,
Al-Ge systems~\cite{MEX}. When considering the 2D-sheets, 
the physical bending of the nanotubes is found to affect only Al-Si but not Al-Ge. 
In fact, while we do not find any significant change
in the calculated band gap between nanotube (3.9 eV) and 2D-sheet (4.0 eV) for the less bent Al-Ge tube,
for the Al-Si tube, due to the smaller radius of curvature,
the calculated band gap is 0.4 eV smaller than for the corresponding 2D-sheet.
For completeness we considered the Al-Si (Al-Ge) 2D-sheets also
by adopting the hybrid functional B3LYP~\cite{B3LYP} which
is reported to account correctly for the chemical reactivity of 
aluminosilicate-based materials~\cite{ZEO-Hybrid}.
In line with previous HF-based calculations for zeolites,
which produced band gaps considerably larger than PW91 results~\cite{Catlow},
the B3LYP band gaps are wider than both the PW91 values and the experimental optical gap
of 3.6 eV~\cite{3}. Specifically, within the adopted triple-zeta basis (see SI),
we found 9.3 eV and 7.42 eV for Al-Si and Al-Ge 2D-sheets respectively.
On these grounds it is reasonable to expect that B3LYP will also overestimate the band gap for the complete nanotubes.

\begin{figure}[htbp]
\begin{center}
\includegraphics[width=\columnwidth]{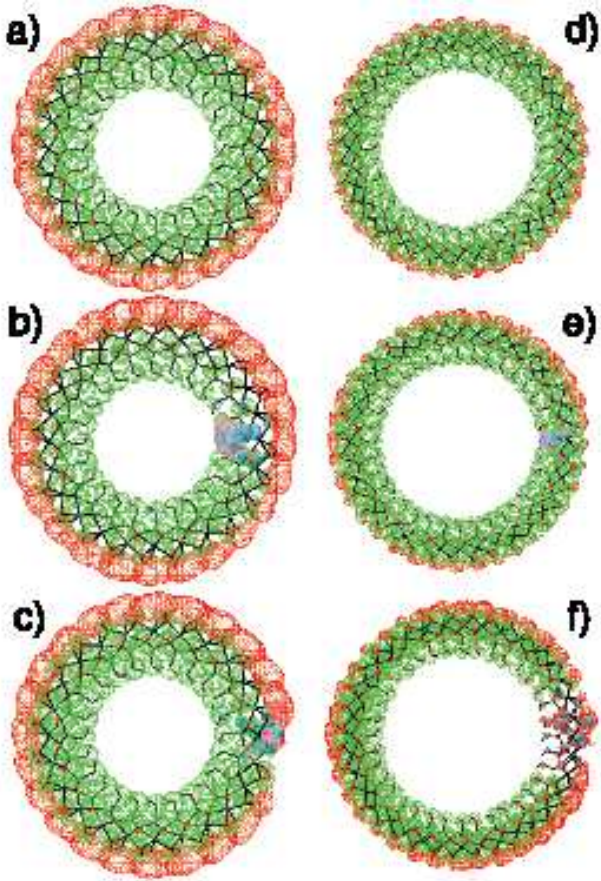}
\end{center}
\caption{Band-decomposed charge density for defect-free Al-Si (a) and Al-Ge (d),
and in the presence of one $\mathrm{OH_{vac}}$ both inside (b,e) and outside (c,f) the nanotube cavity.
The VB (green, $\mathrm{10^{-6} e\AA^{-3}}$) and CB (red, $\mathrm{5\times10^{-7} e\AA^{-3}}$) densities have been
integrated over 0.5 eV from the band onset.
Occupied and empty band-gap defect states are displayed ($\mathrm{5\times10^{-7} e\AA^{-3}}$) in cyan and pink, respectively.}
\label{fig3}
\end{figure}

Fig. 3 shows the band-decomposed charge densities for both the valence band (VB) and conduction band (CB) edges,
calculated by considering electronic states within 0.5 eV from the band onsets.
Interestingly, it emerges that both Al-Si and Al-Ge are characterized by
a neat separation in real space of VB and CB.
The VB edge is in fact localized inside the nanotube cavity,
while the CB edge faces the outer side of the nanotube.
In line with~\cite{MEX}, these findings confirm a strong radial anisotropic
electron affinity with an ensuing enhanced Br{\o}nsted acidity (basicity) for
the outer (inner) tube surfaces as suggested in~\cite{GUSTAF} (see SI for further details). 
In order to assess the effect of OH vacancies on the global electronic 
structure of the tube, one hydroxyl fragment was eliminated both
on the inner, $\mathrm{Al-Si(Ge)-OH_{invac}}$, and on the outer, $\mathrm{Al-Si(Ge)-OH_{outvac}}$, side. 
In analogy with O vacancies on other metal-oxide substrates~\cite{MO-SURF},
the presence of one $\mathrm{OH_{vac}}$ is found to introduce electronic states
in the pristine band gap (see Fig. 2). One $\mathrm{OH_{invac}}$ is found to
create both occupied and unoccupied defect states,
which reduce the actual band-gap to $\mathrm{\sim}$1.8 eV and $\mathrm{\sim}$1.1 eV for both Al-Si and Al-Ge.
This effect thus enhances the conductivity for the global systems substantially.
Both defects are  magnetic, generating a doublet spin-state.
In line with previous results for paramagnetic defects in zeolites~\cite{Catlow},
both for Al-Ge and Al-Si, the occupied (unoccupied) defect states (see Fig. 3 and SI)
are highly localized around the undercoordinated Si (Ge) atom.
Interestingly, while one $\mathrm{OH_{vac}}$ on the outer wall of Al-Si
is also found to create a localized paramagnetic (doublet) spin-state and
to reduce the actual band gap (0.8 eV), the same defect in the Al-Ge analogue 
forms a state just at the CB onset which pins the Fermi level there. 
Consequently, despite the odd total electron count in the system, our DFT energy is minimized by a 
non-magnetic solution with equal occupancy of both up and down spins~\cite{metal} (see Fig. 3 and SI).
We cannot be sure from our calculations whether this behaviour would also be found for a truly isolated defect,
or is an artefact of the periodic defect structure along the tube imposed by our boundary conditions,
but we would expect the binding energy of any localized state to be extremely small.
The calculated energies for both $\mathrm{OH_{outvac}}$ occupied states (Fig. 2) are very close,
suggesting that the reduced (yet still overestimated in PW91~\cite{3}) band gap of Al-Ge is the cause of this metallization.

\begin{figure}[htbp]
\begin{center}
\includegraphics[width=\columnwidth]{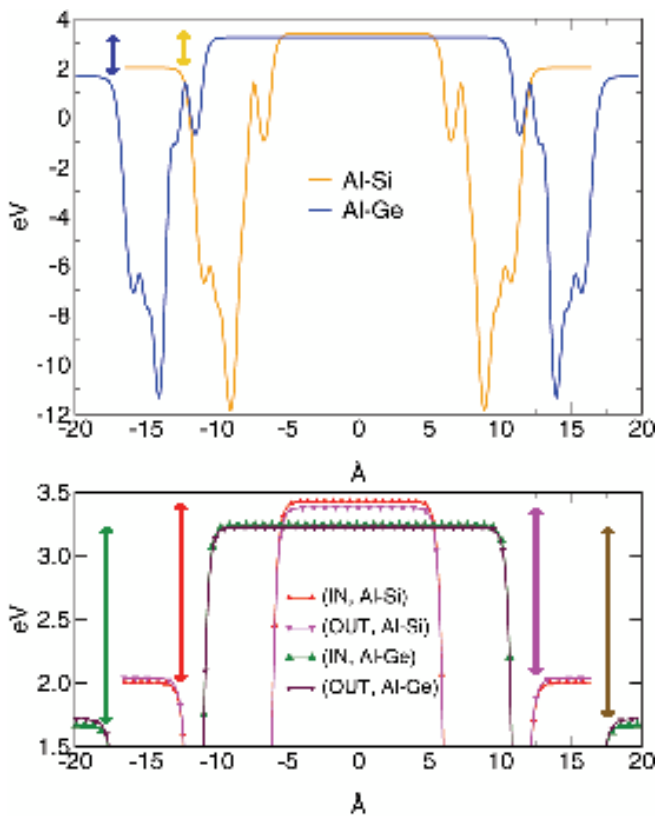}
\end{center}
\caption{(Color online) Top: radial distribution of the averaged electrostatic potential $V$ (see text for details of averaging).
Bottom: radial $V$ shifts as a function of the considered $\mathrm{OH_{vac}}$ defects. 
Vertical arrows highlight the electrostatic potential shifts $\Delta V$ across the tube walls.}
\label{fig4}
\end{figure}

It has been suggested that surface charge properties and chemical reactivity
of Imogolite depend on a delicate (pH-dependent) balance between
local geometrical deformations associated with the tube bending and 
positive (negative) charge accumulation on the outer (inner) surface~\cite{GUSTAF}.
Charge separation across the tube walls is expected to play a fundamental role also
for the suggested use of Al-Si (Al-Ge) as an insulating coating around conducting cores~\cite{3,ZEOM}.
In this respect, it is of utmost importance to assess the changes induced
by $\mathrm{OH_{vac}}$ in terms of both local and global polarizations across the tube section.
Following ~\cite{VDW}, the change ($ \Delta V $) in the (microscopically) averaged
electrostatic potential across the polarized interface can be related to
the electrostatic dipole across the interface itself ($\mu$) as $ \Delta V = e \mu / \epsilon_0$.
On this basis, it is straightforward to evaluate the actual dipole moment
across the tube walls from $\Delta V$ values calculated in DFT.
From the shift in the electrostatic potential (averaged both angularly and along the tube axis, see Fig. 4),
Al-Ge is calculated to possess a larger dipole across the wall (0.676 Debye or
roughly one third of $\mathrm{H_2O}$, 1.85 D) with respect to Al-Si (0.591 D).
In line with~\cite{GUSTAF} (and at odds with~\cite{MEX}) the calculated dipole directions
agree with the positive (negative) charge accumulation at the outer (inner) surface.
Despite its larger dipole, Al-Ge is characterized by lower inner and outer plateaus
of $V$ with respect to Al-Si. This suggests that the net polarization comes from
a delicate balance involving the total electronic distribution across the whole interface,
rather than the atomic charges alone. Owing to the angular symmetry of the tubes,
both Al-Ge and Al-Si do not possess any permanent global dipole moment.
Averaging~\cite{BOTHER} the electrostatic potential over the circular sector
affected by the defect states (see Fig. 3),
it turns out that the introduction of one $\mathrm{OH_{outvac}}$ with the ensuing
electronic charge accumulation on outer wall slightly lowers the corresponding dipole
by 0.021 D (0.028 D) for Al-Si(Ge).
The results are less intuitive (at least with respect to
the defect induced charge accumulation in Fig. 3) for $\mathrm{OH_{invac}}$ defects,
where the corresponding dipole is increased (decreased) by 0.014 D (0.008 D) for Al-Si(Ge),
further suggesting that the total defect-induced redistribution of electron charge
must be taken into account even for the highly localized perturbations found here.
Interestingly, the lack of global polarization is maintained for both Al-Ge and Al-Si
even in the presence of one $\mathrm{OH_{vac}}$.
These findings suggest a capacity of both Al-Ge and Al-Si to counterbalance intrinsically
local charge accumulations and to damp any ensuing global polarization, which could be highly 
relevant for potential applications as insulating components in one-dimensional conducting-semi-conducting hybrid nanodevices.

We gratefully acknowledge support from the Engineering and Physical Sciences Research Council (EPSRC, UK, EP/C541898/1).
WAH acknowledges support from the Royal Society.


\begin{thebibliography}{30}
\bibitem{BOTHER}
By changing the angular sector for the average procedure from $\pm5^{\circ}$ to $\pm30^{\circ}$ degrees
around the defect sites, the reported values were found to be converged within $\pm$0.003 D.
\end{thebibliography}
\end{document}